\documentclass{article}
\usepackage[pdftex]{graphicx}
\usepackage{wrapfig}
\usepackage{amsmath}

\usepackage{orcidlink} 
\hypersetup{
    hidelinks
} 

\PassOptionsToPackage{numbers,sort&compress}{natbib}

\usepackage[preprint]{neurips_2026}

\usepackage[utf8]{inputenc} 
\usepackage[T1]{fontenc}    
\usepackage{hyperref}       
\usepackage{url}            
\usepackage{booktabs}       
\usepackage{amsfonts}       
\usepackage{nicefrac}       
\usepackage{microtype}      
\usepackage{xcolor}         

\title{Exploratory Experience Shapes the Geometry of Predictive Representations}

\author{%
  Kseniia Shilova \orcidlink{0009-0004-2053-7558}\thanks{Corresponding author} \\
  School of Mathematics\\
  Georgia Institute of Technology\\
  Atlanta, GA 30332 \\
  \texttt{kshilova3@gatech.edu} \\
  \And
  Abdelrahman Sharafeldin \orcidlink{0009-0000-2706-0300}\\
  ML@GT \\
  Georgia Institute of Technology \\
  Atlanta, GA 30332 \\
  \texttt{abdo.sharaf@gatech.edu} \\
  \And
  Advay Balakrishnan \\
  College of Computing \\
  Georgia Institute of Technology \\
  Atlanta, GA 30332 \\
  \texttt{abb32@gatech.edu} \\
  \And
  Hannah Choi \orcidlink{0000-0002-8192-1121}\thanks{Corresponding author}\\
  School of Mathematics\\
  Georgia Institute of Technology \\
  Atlanta, GA 30332 \\
  \texttt{hannahch@gatech.edu} \\
}

\begin{document}

\maketitle

\begin{abstract}
Active sensing links behavior and learning through an action-perception loop: actions determine the observations used to update internal predictive models of perception, which subsequently guide the next actions. Predictive-coding frameworks provide a natural way to model this process, since internal representations are continuously updated to predict future observations. Here, we ask how exploratory and exploitative behavioral strategies shape these internal predictive representations. To analyze this relationship, we build an online learning agent in a tree-like maze with a controllable parameter regulating the balance between exploratory and exploitative regimes. The agent updates a predictive-coding-based perception model from experience generated by its own behavior. The model predicts both future maze states and reward probability, allowing the agent to select actions either by expected information gain during exploration or by predicted reward during exploitation. We show that the resulting internal predictive representations depend strongly on the agent’s behavioral regime. Exploratory agents develop representations that are more spatially organized and better preserve the structure of maze transitions in latent space. In contrast, exploitative agents learn less organized representations due to limited exploratory experience. We then train this predictive model on natural trajectories of water-deprived mice navigating the same maze and compare the resulting representations with those learned from agent trajectories. More exploratory mice show representational geometries that closely match those of exploratory agents, whereas mice with more restricted visitation patterns resemble reward-driven, exploitative agents. Together, these findings suggest that exploration enables predictive models to form generalized internal representations by organizing latent space around both spatial location and transition context in artificial agents and animals.
\end{abstract}

\section{Introduction}

Navigating agents learn from the trajectories they generate. Each action determines the next observation, and each observation provides evidence for updating an internal model of the environment. This closed action-perception loop is central to active sensing: behavior shapes what the animal observes, and therefore what it can learn \citep{friston2010free, little2013learning, yang2016active}. The same idea is important for artificial embodied agents \citep{linson2018active}, where the policy used to sample the environment determines the data available for learning.

In spatial navigation, different behavioral strategies expose the learner to different parts of the environment. While exploratory sampling uncovers the broader transition structure of a space, reward-driven behavior often reinforces a narrow set of reliable paths, sacrificing variability for efficiency. This transforms the exploration-exploitation balance into a representational question: how does the way an agent samples an environment shape the internal model it learns?

This question is closely related to biological theories of spatial representation. Cognitive-map theories propose that animals form internal representations of space that support flexible navigation \citep{tolman1948cognitive, okeefe1978hippocampus}. Recent computational work suggests that map-like representations can emerge from sequential or predictive learning: models trained to predict future observations or transitions can develop latent states that reflect spatial structure, trajectory context, or hidden environmental state \citep{george2021clone, raju2022space, levenstein2024sequential}. These results show that spatial structure can emerge from predictive objectives, but they leave open how the behavioral policy generating the training trajectories affects the structure of the learned map.

We study this question by isolating the effect of behavioral sampling in an online learning agent. Our agent navigates a binary-tree maze motivated by a mouse labyrinth task \citep{rosenberg2021mice} in which mice can navigate to a fixed reward site and explore the maze. The agent uses a single predictive-coding-based perception model to learn from its own experience and guide future actions. Given the current state and intended action, the model predicts both the next position within the maze and the probability of receiving a reward there. These predictions support action selection based on two behavioral modes: an exploratory mode, where actions are selected to increase expected information gain, and a reward-driven mode, where actions follow a value map derived from predicted rewards. By varying the probability of entering the reward-driven mode, we generate agents with different exploration-exploitation balances while keeping the predictive model and learning objective fixed.

We find that behavioral strategy defines not only the ability of the agent to reach reward, but also the organization of the learned predictive space. Pure exploration produces accurate transition models and spatially organized latent geometry with smooth trajectories through the maze. Adding reward-driven behavior improves reward seeking when it is balanced with continued exploration, but it also narrows the agent's experience and weakens the organization of the learned latent map. Training the model on real mouse trajectories reveals a similar divergence in behavioral modes: whereas mice that engage in extensive exploration develop spatially organized latent geometries, those driven by rewards exhibit disrupted internal representations, similar to the distinct representation geometries in exploratory and reward-driven agents.

\section{Related work}
\label{sec:related_work}

\paragraph{Predictive coding, active sensing, and behavioral regimes.}
Predictive coding frames perception as iterative prediction and prediction-error correction \citep{rao1999predictive, friston2005theory}. In active sensing, this process is embedded in a closed action-perception loop: actions determine which observations are sampled, and those observations update the agent's internal model \citep{yang2016active}. This creates a direct link between behavior and learning, because the policy used to sample the environment determines the data on which the predictive model is trained. Recent models have used this principle to combine predictive coding with uncertainty minimization, allowing agents to select actions by expected information gain and learn transition structure or active visual representations \citep{sharafeldin2024active, butko2010infomax, rao2024active}. Active inference provides a related account in which action selection can combine epistemic value with expected preferred outcomes, linking information seeking with goal-directed behavior \citep{friston2012perceptions, tschantz2020reinforcement}. 

Natural behavior, however, is rarely purely epistemic or purely reward-driven. Animals dynamically shift between behavioral strategies, and inverse reinforcement learning methods can infer changing intentions or reward functions from long behavioral trajectories \citep{ashwood2022mice,zhu2024multiintention,ke2025swirl}. These lines of work establish that information seeking, reward seeking, and strategy switching are closely related. Our model brings these ideas together in a controlled setting: a single perception model based on predictive coding supports both information-driven exploration and reward-driven action selection, allowing us to vary the sampling regime while keeping the underlying predictive model fixed.

\paragraph{Computational predictive maps.}
Cognitive-map theories propose that animals form internal representations of space that support flexible navigation \citep{tolman1948cognitive, okeefe1978hippocampus}. A growing body of computational work suggests that such map-like representations can emerge from predictive or sequence-learning objectives. In this view, spatial representations need not be explicitly defined; they can arise when a model must predict future states, disambiguate aliased observations, or represent trajectory context. Successor representations formalize this idea by encoding future state occupancy under a policy \citep{dayan1993, stachenfeld2017hippocampus}, while clone-structured cognitive graph models show how latent states can recover spatial topology from sequential experience \citep{george2021clone}. Predictive models trained on navigation sequences can also develop spatially tuned units and low-dimensional latent manifolds that reflect position, trajectory structure, and context \citep{recanatesi2021predictive, raju2022space, banino2018, whittington2020tolman, wang2025time, cone2024latent}. Related hippocampal models link predictive sequence learning to CA3-CA1 dynamics and show that recurrent predictive networks can produce cognitive-map-like structures and trajectories resembling replays \citep{chen2024predictive, levenstein2024sequential}.

Taken together, these studies suggest that the latent states of predictive models can form map-like representations. However, much of this work focuses on the representations that emerge from a given predictive objective or a set of training trajectories. Less is known about how the behavioral policy that generates those trajectories changes the resulting latent geometry. We extend the predictive-map framework by treating behavioral sampling as the experimental variable. Specifically, we ask whether the same predictive objective produces map-like geometries when the agent explores broadly versus when it repeatedly samples reward-directed paths.

\paragraph{Experimental representational geometry in navigation.}
Growing experimental evidence suggests that navigational representations are best characterized as experience-dependent population geometries, rather than mere ensembles of single-unit place fields. Hippocampal assemblies reorganize around newly learned goals, and their reactivation predicts spatial memory performance \citep{dupret2010reorganization}. More broadly, hippocampal population geometry changes with experience: CA1 spatial representations can expand with exploration time in a hyperbolic geometry, and latent learning can gradually transform high-dimensional CA1 activity into a low-dimensional manifold resembling the explored environment \citep{zhang2022hyperbolic, guo2024latent}. Population-level analyses further show that hippocampal and related circuits organize task variables geometrically, including spatial position, evidence, speed, future path, and reward-related events \citep{nieh2021geometry, nakai2024distinct}.

These studies suggest that navigation should be studied not only through the decoding of position or reward, but also through the population-level representational geometry. At the same time, experimental data alone cannot separate the effects of learning objective, task structure, and behavioral sampling. Our approach uses a controlled predictive model to isolate the effect of sampling strategy, and analyzes mouse navigational trajectories based on the latent representational space. This allows us to probe whether exploratory behavioral experience is associated with spatially organized predictive geometry in both artificial agents and models trained on animal behavior.

\begin{figure}[t]
  \centering
  \includegraphics[width=0.8\linewidth]{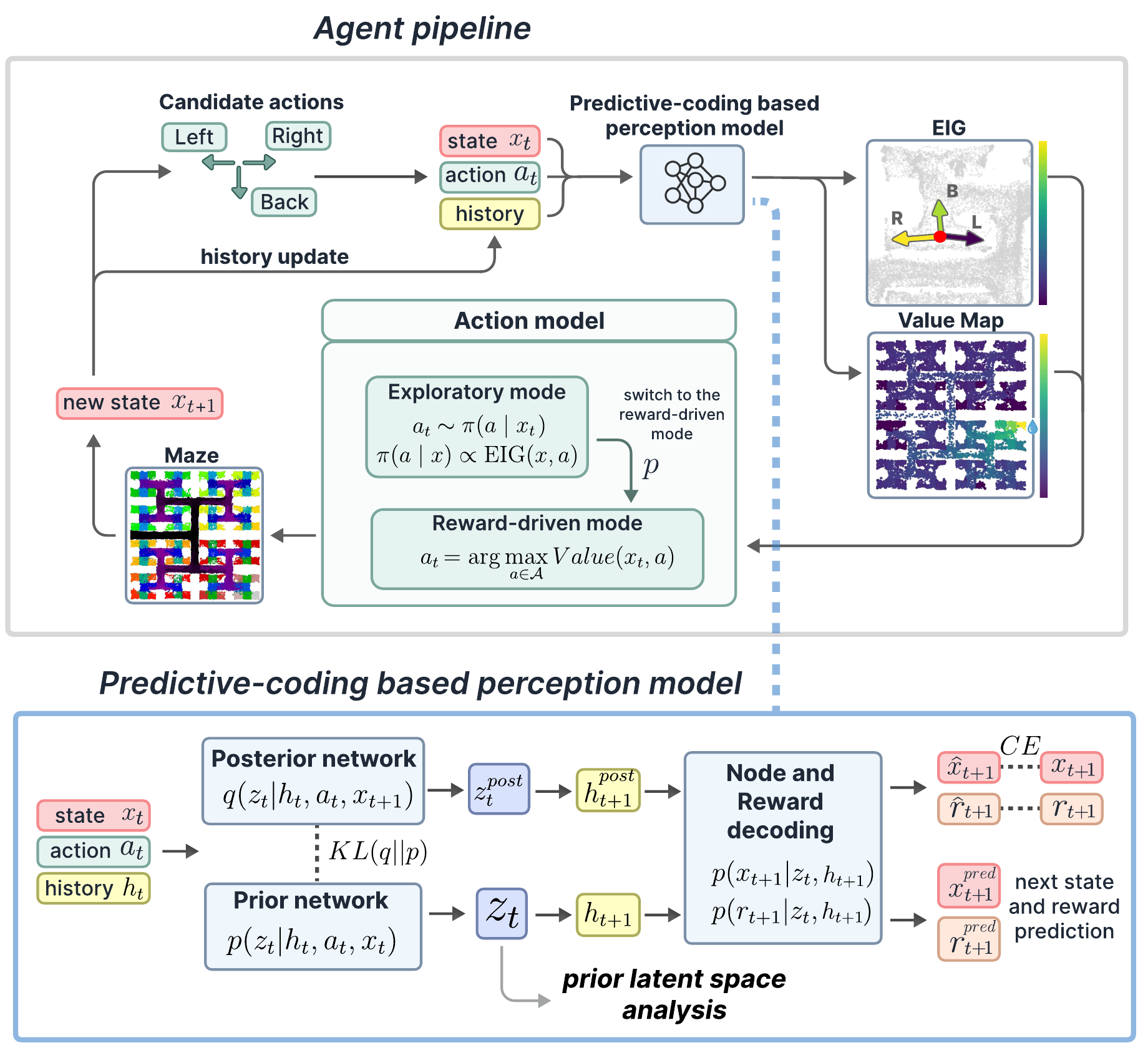}
  \caption{\textbf{Predictive-coding agent with exploration-exploitation switching.}
\textbf{Top:} Overview of the action-perception loop. At each step, the agent evaluates locally valid actions using its current predictive model. In the exploratory regime, actions are sampled according to expected information gain (EIG); in the reward-driven regime, actions are selected using a value map constructed from learned reward predictions and the transition graph observed through experience. After acting, the agent observes the next maze state, updates its recurrent state, and continues the loop.
\textbf{Bottom:} Action-conditioned predictive model. Before observing the transition outcome, the prior network predicts a latent transition state ($z_t$) from the current recurrent state ($h_t$), current maze node ($x_t$), and intended action ($a_t$). After the next node ($x_{t+1}$) is observed, the posterior network infers a posterior latent state ($z_t^{post}$). The latent state updates the recurrent history and is used to decode both the next node and reward probability. The prior latent state is used for latent-space analyses.}
  \label{fig:fig1}
\end{figure}

\section{Task and modeling framework}

We model navigation in the binary-tree labyrinth introduced by \citet{rosenberg2021mice}. This experimental paradigm, in which water-deprived mice learn to navigate to a single rewarded leaf node while maintaining broad exploration, provides an ideal framework for our analysis: it integrates reward-directed navigation with persistent exploratory behavior, allowing us to examine how these dual objectives shape representational geometry.

Following the same abstract task structure, the environment is modeled as a deterministic binary tree with $N=127$ nodes (Appendix Fig.~\ref{fig:app1}A). Each bout begins at the root node. At time $t$, the agent occupies node $x_t$ and chooses one locally valid action
\[
a_t \in \{\mathrm{left},\mathrm{right},\mathrm{back}\}.
\]
The action deterministically moves the agent to the next node $x_{t+1}$. A sparse reward is delivered only when the agent reaches a fixed reward node $x^\star$,
\[
r_{t+1} = \mathbb{I}[x_{t+1}=x^\star].
\]
The agent learns online from trajectories generated by its own policy, so the behavioral mode directly determines the data available for learning.

The agent is built around an action-conditioned predictive-coding model, shown in Fig.~\ref{fig:fig1}. We use ``predictive-coding'' to refer to the computational objective of minimizing the separation between an action-conditioned prior prediction and a posterior state inferred after observing the transition. The model maintains a recurrent hidden state $h_t$ summarizing previous transitions. Before observing the outcome of action $a_t$, the prior network predicts a latent transition state $z_t$ based on:
\[
p_\theta(z_t \mid h_t,a_t,x_t).
\]
After the transition is observed, the posterior network infers a latent state using the realized next node, following:
\[
q_\phi(z_t \mid h_t,a_t,x_{t+1}).
\]
The recurrent state is updated as
\[
h_{t+1}=f_\psi(h_t,a_t,z_t),
\]
where $f_\psi$ is a GRU update function.

The model decodes both the next node $x_{t+1}$ and reward probability $r_{t+1}$, following: 
\[
p_\theta(x_{t+1}\mid z_t,h_{t+1})
\qquad \mbox{and} \qquad
p_\theta(r_{t+1}=1\mid z_t,h_{t+1}),
\]
respectively. The decoder is shared between prior and posterior computational paths. During training, the posterior latent state \(z_t^{post}\) updates the recurrent history \(h_{t+1}^{post}\), and the pair \((z_t^{post}, h_{t+1}^{post})\) is decoded to predict the observed next node and reward \((\hat{x}_{t+1}, \hat{r}_{t+1})\). The next-state and reward cross-entropy losses are computed from these posterior-path predictions. During action selection, the model instead uses the prior latent and recurrent states \((z_t, h_{t+1})\) to predict the next state and reward (\(x_{t+1}^{pred}, r_{t+1}^{pred}\)). These prior-path predictions are used for computing expected information gain (EIG), reward-map construction, and latent-space analyses.

The model is trained online with a predictive-coding objective,
\[
\mathcal{L}
=
\mathcal{L}_x
+
\beta \mathcal{L}_{\mathrm{KL}}
+
\lambda_r \mathcal{L}_r
+
\lambda_s \mathcal{L}_{\mathrm{sparse}}.
\]
Here $\mathcal{L}_x$ is the categorical cross-entropy for next-node prediction and $\mathcal{L}_r$ is the binary cross-entropy for reward prediction. $\mathcal{L}_{\mathrm{KL}}$ aligns posterior and prior latent distributions,
\[
\mathcal{L}_{\mathrm{KL}}
=
D_{\mathrm{KL}}
\left[
q_\phi(z_t\mid h_t,a_t,x_{t+1})
\;\|\;
p_\theta(z_t\mid h_t,a_t,x_t)
\right],
\]
and $\mathcal{L}_{\mathrm{sparse}}$ regularizes the posterior latent mean.

Action selection has two regimes. In the exploratory regime, the agent scores locally valid actions by expected information gain (EIG). Because the latent state represents the model's belief about the upcoming transition, reducing uncertainty in this latent state provides a natural exploration objective. We calculate this information gain as the expected reduction in latent entropy after observing the next node,
\[
\mathrm{EIG}(x_t,a)
=
H[p_\theta(z_t\mid h_t,a,x_t)]
-
\mathbb{E}_{\hat{x}_{t+1} \sim p_\theta(x_{t+1}\mid z_t,h_{t+1})}
H[q_\phi(z_t\mid h_t,a,\hat{x}_{t+1})].
\]
In the reward-driven regime, actions are selected using a value map built from the model's recent reward predictions. Specifically, predicted reward probabilities are averaged over recently visited nodes to form a learned reward map $\hat{R}(x)$, and values $V(x)$ are propagated over the transition graph observed by the agent (Appendix Fig.~\ref{fig:app2}A).
\[
V(x)
\leftarrow
\hat{R}(x)
+
\gamma T
\log
\sum_{x' \in \mathcal{N}_{\mathrm{obs}}(x)}
\exp\left(\frac{V(x')}{T}\right),
\]
where $\mathcal{N}_{\mathrm{obs}}(x)$ denotes the empirically observed successor nodes of $x$, $\gamma$ is a discount factor, and $T$ is the softmax temperature. The reward-driven policy then chooses the valid action leading to the neighboring node with highest learned value.

The exploration-exploitation balance is controlled by a single switching probability $p$. At each exploratory step, the agent enters reward-driven mode with probability $p$. Once reward-driven mode is active, the agent follows the value-map policy until it reaches the reward node; after reward delivery, it is forced back into exploratory behavior for a fixed number of steps (\(n=7\)). When $p=0$, the agent is purely exploratory, while larger values of $p$ produce increasingly reward-driven agents. Importantly, all agents share the same environment, architecture, loss function, reward location, and online learning procedure.

\section{Exploration–exploitation balance controls behavioral sampling and learning}
\label{sec:behavior}

\begin{figure}[t]
  \centering
  \includegraphics[width=0.95\linewidth]{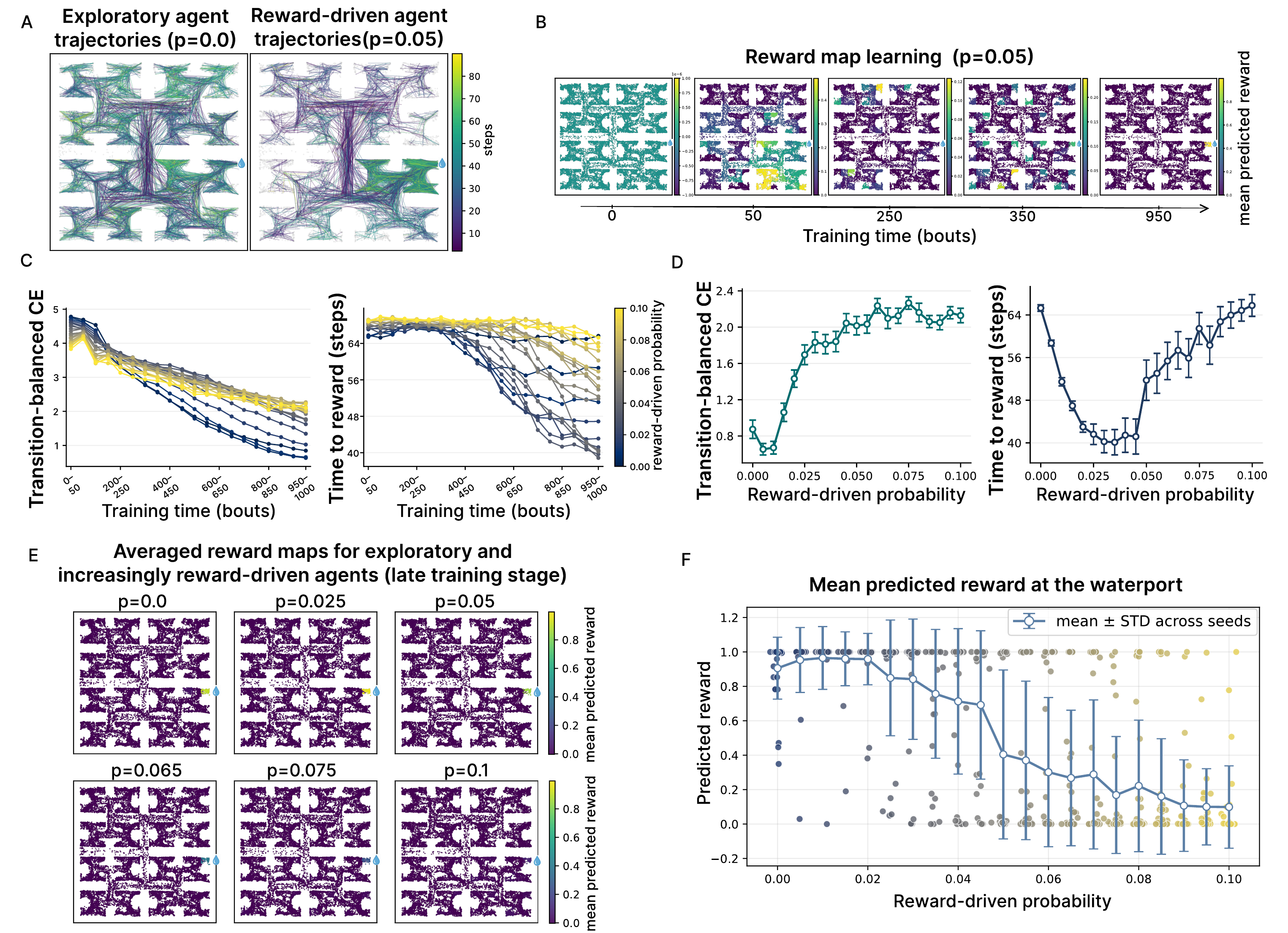}
  \caption{\textbf{Behavioral consequences of exploration-exploitation balance.}
\textbf{A:} Example trajectories of an exploratory agent and a more reward-driven agent. The reward-driven agent visits the water port more often, whereas the exploratory agent samples a broader range of maze branches.
\textbf{B:} Evolution of the learned reward map over training. The reward map is constructed from recent experience by averaging predicted reward at visited locations. Over learning, the map gradually concentrates around the rewarded end node.
\textbf{C:} Transition-balanced cross-entropy (CE; left) and time to reward (right) over learning for different reward-driven switching probabilities $p$. More exploratory agents learn the transition structure faster and achieve lower CE. Time to reward depends non-monotonically on $p$: agents learn the reward location more efficiently and reach the reward faster when $p$ initially increases from 0, but the efficiency degrades when agents are highly reward-driven.
\textbf{D:} Final-stage transition-balanced CE (left) and time to reward (right) as a function of $p$. Transition learning degrades as reward-driven behavior increases, while reward-seeking performance is best at an intermediate exploration-exploitation balance ($p \approx 0.025$-$0.05$).
\textbf{E:} Mean learned reward maps at the final learning stage. Largely exploratory agents with a low but non-zero probability of reward-driven behavior show stronger localization of reward around the true water-port, while more reward-driven agents learn weaker or less accurate reward maps.
\textbf{F:} Mean predicted reward at the true water-port node during the final learning stage, averaged across 30 seeds. Exploratory and moderately reward-driven agents reliably identify the rewarded node, whereas strongly reward-driven agents often fail to do so.
}
  \label{fig:fig2}
\end{figure}

We next investigated how the probability of entering reward-driven mode affects learning and navigation. We trained agents across multiple values of $p$, the probability of entering reward-driven mode at an exploratory step.

This manipulation changes the distribution of trajectories experienced by the agent (Fig.~\ref{fig:fig2}A). Exploratory agents visited a larger fraction of maze branches and depths. More reward-driven agents concentrated visits along paths favored by the learned value map, producing more frequent trajectories toward the rewarded region but less coverage of the full tree.

This sampling difference created a tradeoff between transition learning and reward seeking. Transition learning degraded as reward-driven behavior increased: agents with lower $p$ achieved lower transition-balanced cross-entropy (CE) throughout training, while agents with higher $p$ learned the transition structure more slowly and had higher final CE (Fig.~\ref{fig:fig2}C,D). This is consistent with their more restricted experience, which provides fewer distinct transitions for updating the predictive model.

Reward seeking showed a non-monotonic dependence on $p$. Purely exploratory agents learned accurate transition models, but they did not preferentially return to the reward location, so their time to reward remained high. Agents with intermediate reward-driven probability reached the reward more efficiently because they retained enough exploration to learn useful transition and reward predictions while also exploiting the emerging value map. When reward-driven behavior became too frequent, performance worsened again: these agents exploited incomplete value maps, and their restricted sampling limited the experience needed to correct them (Fig.~\ref{fig:fig2}C,D).

The learned reward maps explain this tradeoff. During training, predicted reward gradually becomes localized around the true water-port and then propagates through the observed transition graph (Fig.~\ref{fig:fig2}B). This process requires sufficient coverage of terminal nodes and transitions. Consistent with this interpretation, exploratory and moderately reward-driven agents assign higher predicted reward to the true water-port node, whereas agents with larger $p$ often learn ``weaker" and less accurate reward maps (Fig.~\ref{fig:fig2}E,F).

Taken together, these results show that reward-guided behavior is effective only when combined with sufficient exploration. Exploration improves the predictive world model and provides the coverage needed to form an accurate reward map; exploitation becomes beneficial once that map contains enough structure to guide navigation. Thus, transition learning degrades as behavior becomes more exploitative. Interestingly, reward-seeking performance peaks when exploration and exploitation are balanced, surpassing the performance seen in purely exploitative settings.

\section{Exploratory sampling organizes predictive latent geometry}
\label{sec:latent_geometry}

\begin{figure}[t]
  \centering
  \includegraphics[width=0.95\linewidth]{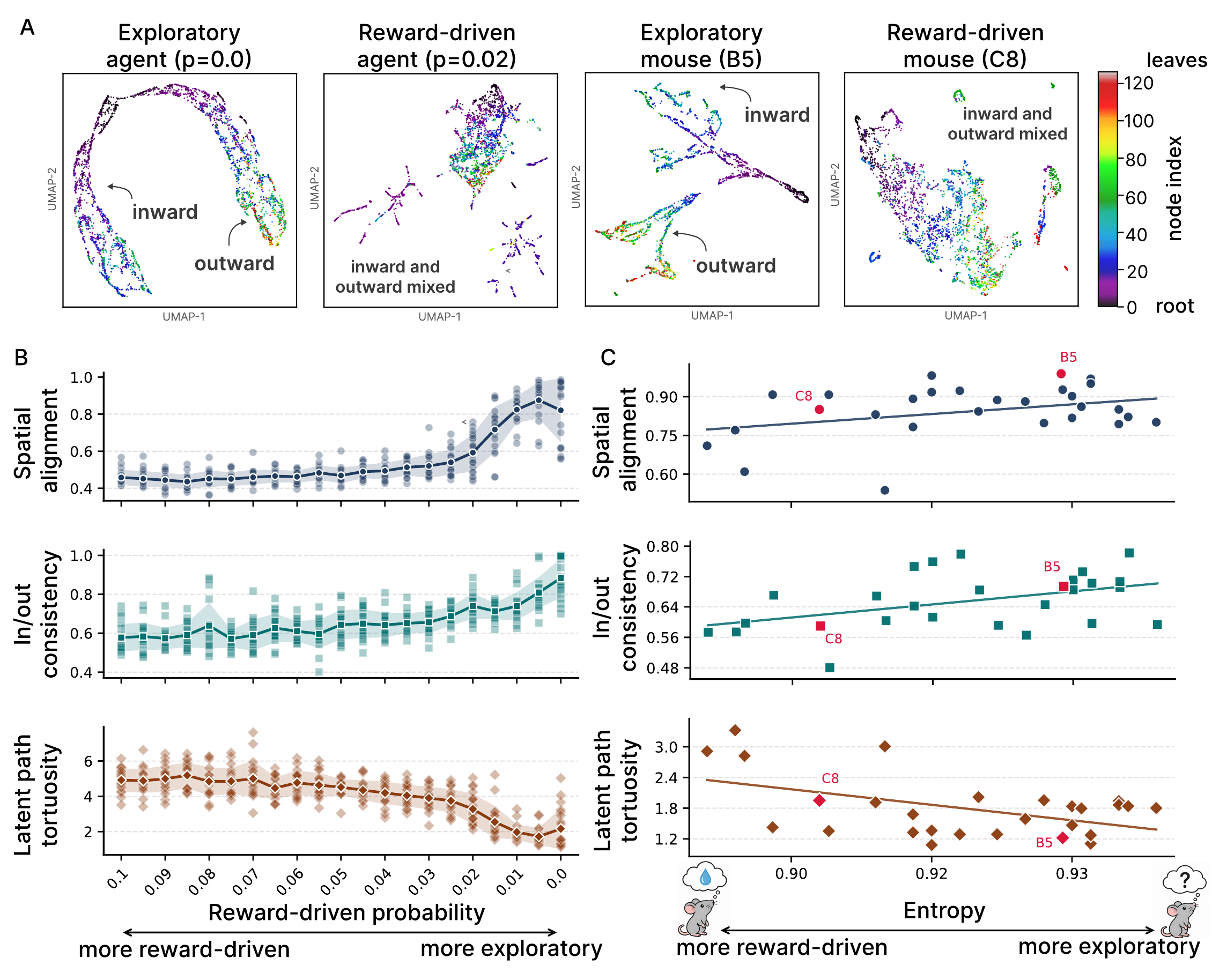}
  \caption{\textbf{Exploration shapes the geometry of predictive latent representations in agents and mice.}
\textbf{A:} UMAP visualizations of prior latent states for exploratory and reward-driven agents, and for two example mice. Exploratory agents develop depth-aligned, branched latent spaces with clearer separation between outward (root-to-leaf) and inward (leaf-to-root) transitions. More reward-driven agents show less organized latent spaces with disrupted branch structure and reduced spatial alignment. Similar differences are observed between the exploratory mouse B5 and the less exploratory mouse C8.
\textbf{B:} Quantitative measures of latent geometry for agents, averaged across 30 seeds. We measure spatial alignment, transition consistency, and trajectory tortuosity. Exploratory agents show stronger depth alignment, stronger local grouping of transition direction, and smoother latent trajectories.
\textbf{C:} The same latent-space metrics computed for mice. Red points mark mice B5 and C8, whose UMAPs are shown in panel A. More exploratory mice show latent geometries closer to exploratory agents, whereas less exploratory mice resemble mildly reward-driven agents.}
  \label{fig:fig3}
\end{figure}

We next asked whether the behavioral differences induced by exploration and exploitation are reflected in the internal geometry of the predictive model. We focus on the prior latent state,
\[
z_t^p \sim p_\theta(z_t \mid h_t,a_t,x_t),
\]
that represents the model's action-conditioned prediction of the upcoming transition. It is therefore the appropriate object for asking how experience shapes the model's internal predictive map.

The transition-learning results in Fig.~\ref{fig:fig2} suggest that exploratory agents acquire a more accurate model of the maze. We asked whether this improvement is accompanied by a more organized latent space. Fig.~\ref{fig:fig3}A shows UMAP visualizations of prior latent states collected during navigation. Exploratory agents form a branched latent geometry that reflects the hierarchical structure of the binary tree: nearby latent states correspond to related transitions (same maze depth), outward and inward movements are clearly separated, and trajectories through the maze appear as smoother paths in latent space. In contrast, more reward-driven agents produce compact and less organized representations, without a clear separation between outward and inward transitions.

These geometric differences reflect the different sampling distributions induced by each behavioral regime. Exploratory agents sample many branches and depths of the maze, repeatedly exposing the model to shared structural motifs such as movement away from the root, movement back toward the root, and transitions at similar depths. This broad sampling encourages a latent code organized around transition structure. Reward-driven agents, by contrast, repeatedly sample trajectories favored by the current value map. This can support prediction on frequently visited paths, but provides weaker pressure to organize the full transition structure of the maze.

To quantify these differences, we computed three metrics to characterize geometric structures of the original latent space (not on the UMAP projections). First, \emph{spatial alignment} ($\rho_{\mathrm{depth}}$) measures whether latent trajectories preserve the ordering of maze depth. For each monotonic inward or outward trajectory segment of length $L$ with latent states $z_{1:L}$ and maze depths $d_{1:L}$, we project the latent states onto their first principal component and compute
\[
\rho_{\mathrm{depth}}
=
\left|
\mathrm{Spearman}
\left(
\mathrm{PC}_1(z_{1:L}), d_{1:L}
\right)
\right|.
\]
High values indicate that movement through the maze hierarchy corresponds to an ordered direction in latent space.

Second, \emph{transition consistency} ($C_i$) measures whether inward and outward transitions are locally grouped. For each latent transition state $z_i$, we find its $k$ nearest neighboring transition states and compute the fraction of them with the same transition direction:
\[
C_i
=
\frac{1}{k}
\sum_{j\in \mathcal{N}_k(i)}
\mathbb{I}
\{
\mathrm{dir}(j)=\mathrm{dir}(i)
\},
\qquad
\mathrm{dir}\in\{\mathrm{in},\mathrm{out}\}.
\]
High values indicate that inward and outward transitions occupy locally distinct regions of latent space rather than being mixed.

Third, \emph{trajectory tortuosity} ($\tau$) measures whether continuous paths in the maze remain smooth in latent space. For each monotonic trajectory segment, we compute
\[
\tau
=
\frac{
\sum_{\ell=1}^{L-1}
\|z_{\ell+1}-z_\ell\|
}{
\|z_L-z_1\|
}.
\]
Lower values indicate that the latent trajectory follows a more direct path between its endpoints.

Together, these metrics test complementary aspects of predictive geometry: whether latent trajectories are ordered by maze depth, whether transition direction is locally organized, and whether paths through the maze are smooth. Exploratory agents score better on all three measures (Fig.~\ref{fig:fig3}B). Thus, exploration improves not only predictive accuracy, but also the organization of the learned representation. Because the latent geometry is not directly optimized, these differences arise indirectly from the trajectories used to train the same predictive objective.

\section{Mouse trajectories exhibit similar relationships between latent geometry and sampling strategy}
\label{sec:mouse_comparison}

The agent results show that experimentally changing the exploration-exploitation balance alters the geometry of the learned predictive representation. We next asked whether a similar relationship is present in real animals navigating the same binary-tree maze. To test this, we trained the same predictive model on mouse trajectory data from the labyrinth task of \citet{rosenberg2021mice} and applied the latent-space analyses used for the agents.

For each mouse, we trained a separate perception model on that animal's sequence of bouts, preserving the temporal order of experience. The architecture, objective, and hyperparameters were kept fixed across animals and agents. To make the agent and mouse analyses comparable, agent bouts were sampled with approximately matched bout lengths and a similar total amount of behavioral experience. This reduces the possibility that differences in latent geometry are driven only by data quantity or model specification, rather than by differences in the structure of the sampled trajectories.

To summarize each animal's behavioral sampling pattern, we computed normalized visitation entropy,
\[
E_{\mathrm{mouse}}
=
\frac{-\sum_i p_i \log p_i}{\log N},
\]
where $p_i$ is the fraction of visits to node $i$ and $N=127$ is the total number of maze nodes. Higher values indicate more distributed sampling of the maze, while lower values indicate behavior concentrated on a smaller subset of nodes. We use this quantity as a simple measure of exploratory behavior. Example trajectories from mice with high and low visitation entropy are shown in Appendix Fig.~\ref{fig:app1}B.

Normalized visitation entropy varied across mice, indicating that some animals sampled the maze broadly whereas others concentrated their visits on a smaller subset of nodes. Fig.~\ref{fig:fig3}A shows two representative examples. The more exploratory mouse B5 produces a branched latent geometry with clearer depth organization and separation of inward and outward transitions, resembling the geometry of exploratory agents. The less exploratory mouse C8 produces a more compact representation without clear branching and less smooth trajectories, qualitatively closer to the geometry observed in more reward-driven agents. Additional visualizations of the inward and outward paths for these mice are shown in Appendix Fig.~\ref{fig:app1}C.

We quantified this relationship using the same three metrics as in the agent analysis: spatial alignment, transition consistency, and trajectory tortuosity. As before, all metrics were computed in the original high-dimensional prior latent space. Mice with higher visitation entropy showed stronger depth alignment, stronger local consistency of inward and outward transition direction, and lower trajectory tortuosity (Fig.~\ref{fig:fig3}C). Mice with lower visitation entropy showed weaker organization along these dimensions.
Because mouse behavior is observational rather than experimentally manipulated here, these results should be interpreted as a geometry-behavior association rather than direct causal evidence. Nevertheless, the association is consistent with the causal pattern observed in agents.

These results suggest that the geometry-behavior relationship observed in agents is not specific to the synthetic behavioral-mode-switching policy. The same pattern in mice supports the interpretation that exploration shapes not only what the model learns about the environment, but also how that knowledge is organized internally.

\section{Discussion}

Our results suggest that exploration may help organize predictive maps by exposing the learner to broader transition structure, while more restricted reward-directed behavior provides a narrower basis for generalization. However, several limitations remain. Our behavioral policy captures only a simplified exploration-exploitation tradeoff. Animals likely combine multiple drives, including novelty, uncertainty, memory, fatigue, and changing motivational state. In addition, the mouse analysis is based only on behavioral trajectories, and we do not compare predictive representations to simultaneously recorded neural activity. We therefore interpret the mouse results as a model-based prediction: exploratory experience may be associated with similar latent geometry-behavioral mode relationships in neural population activity during navigation.

Future work could extend this framework in two complementary directions. First, richer behavioral datasets with visual sensory input could test whether exploratory experience supports generalization across sensory observations, not only across abstract maze states. This would allow us to ask whether broad sampling organizes predictive representations of visually grounded locations. Second, neural recordings during navigation could be used to test whether the model-predicted relationship between latent geometry and behavioral mode is reflected in neural population activity. For example, such data would make it possible to study how representational organization forms over learning and how it relates to known remapping and learning-related changes in hippocampal and related circuits. Consistent with this direction, Appendix B shows that many individual units develop spatially localized tuning in both the recurrent state \(h_t\), which accumulates transition history, and the latent state \(z_t\), which represents the action-conditioned predictive state. This suggests that population-level predictive geometry may coexist with more classical unit-level spatial tuning. It also provides a way to study how single-unit tuning redistributes across behaviorally important paths or locations over learning, linking experimentally observed effects to the computational dynamics of predictive perception. 

\subsection*{Code Availability}
All our code for model training, simulation, and data analysis is available at https://github.com/HChoiLab/Exploratory-Predictive-Representations.

{
\small
\bibliographystyle{unsrtnat}
\bibliography{references}
}

\appendix

\section{Methods}
\label{app:model_training}

\subsection{Maze environment}

\begin{figure}[t]
  \centering
  \includegraphics[width=1\linewidth]{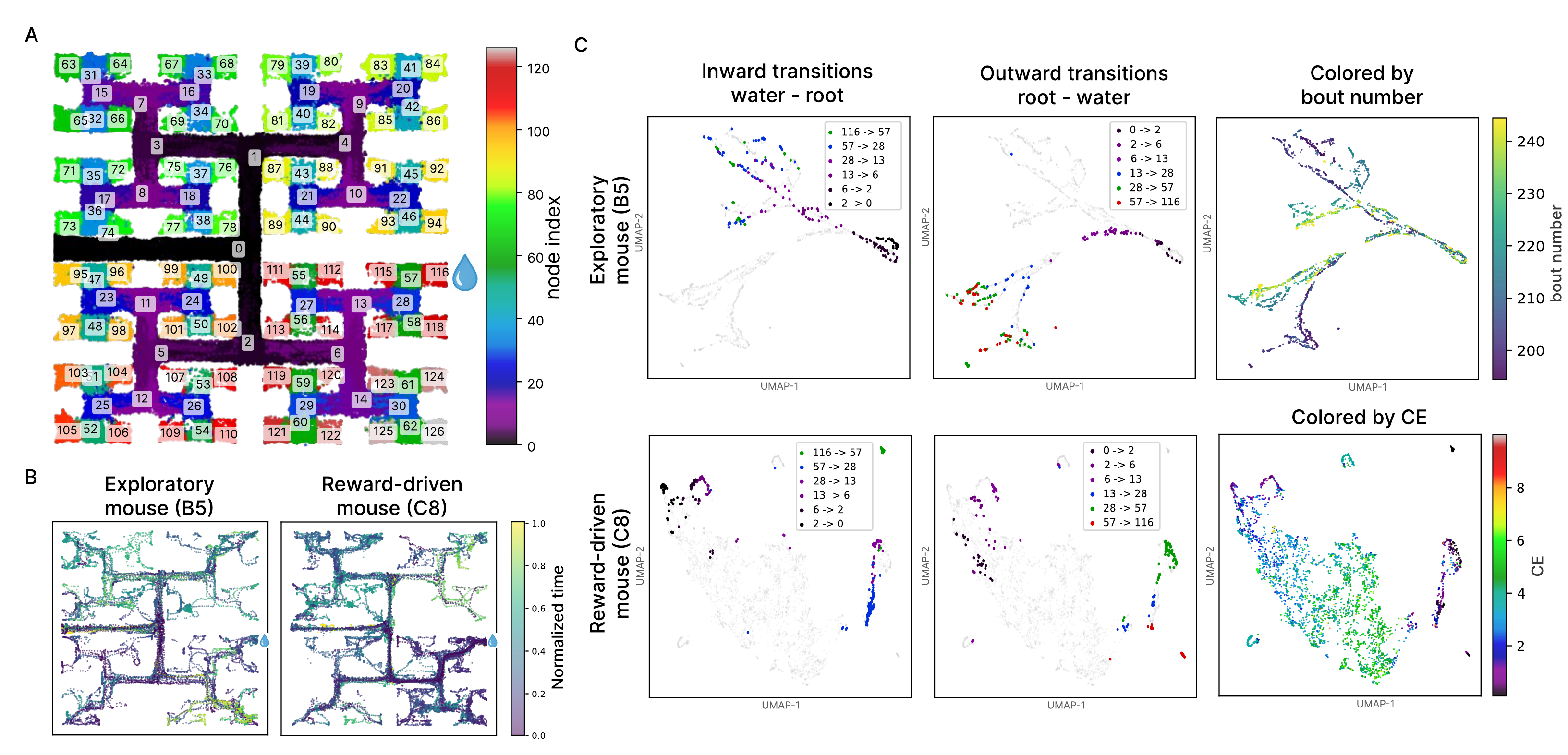}
  \caption{
\textbf{Additional task and mouse-trajectory visualizations.}
\textbf{A:} Binary-tree maze used in the task. Nodes are colored and labeled by index; node 116 is the fixed water-port node.
\textbf{B:} Example trajectories from exploratory mouse B5 and more reward-focused mouse C8. Color indicates normalized step within the plotted bouts. C8 repeatedly follows trajectories toward the water-port, whereas B5 samples the maze more broadly.
\textbf{C:} UMAP embeddings of prior latent states for B5 and C8. Left and middle columns highlight transitions along the inward path from water-port to root and the outward path from root to water-port, respectively. In the right panels, B5 is colored by bout index, revealing learning-related subbranches in the latent representation, and C8 is colored by transition cross-entropy, showing lower prediction error along the frequently visited water-port path.}
  \label{fig:app1}
\end{figure}

The maze is represented as a full binary tree with $N=127$ nodes as shown in Fig.~\ref{fig:app1}A. The valid local actions are \textit{left}, \textit{right}, and \textit{back}. Internal branch nodes have three possible actions, the root has no \textit{back} action, and leaf nodes have only the \textit{back} action. Each bout starts at the root node. At time $t$, the agent occupies node $x_t$, selects a locally valid action $a_t$, and deterministically transitions to $x_{t+1}$. A sparse reward is delivered only when the next node is the fixed water-port node $x^\star$:
\[
r_{t+1}=\mathbb{I}[x_{t+1}=x^\star].
\]
The reward node is fixed throughout training and corresponds to the water-port location in the maze used in \citet{rosenberg2021mice}.

The abstract maze matches the logical binary-tree structure used in the mouse labyrinth task of \citet{rosenberg2021mice}. In both the agent and mouse analyses, behavior is represented as a sequence of node transitions through this tree. This node-level abstraction allows us to train the same predictive model on either self-generated agent trajectories or observed mouse trajectories.

\subsection{Predictive model}

The model maintains a recurrent hidden state $h_t$ ($64$-dimensional) and a stochastic latent transition state $z_t$ ($16$-dimensional). The recurrent state summarizes the history of previous transitions up to the current node. For each transition, a learned prior network maps the current hidden state, current node, and action to the parameters of a diagonal Gaussian distribution:
\[
p_\theta(z_t\mid h_t,a_t,x_t)
=
\mathcal{N}
\left(
\mu^p_\theta(h_t,a_t,x_t),
\operatorname{diag}\left[(\sigma^p_\theta(h_t,a_t,x_t))^2\right]
\right).
\]
After the next node is observed, a learned posterior network maps the hidden state, action, and realized next node to a diagonal Gaussian posterior:
\[
q_\phi(z_t\mid h_t,a_t,x_{t+1})
=
\mathcal{N}
\left(
\mu^q_\phi(h_t,a_t,x_{t+1}),
\operatorname{diag}\left[(\sigma^q_\phi(h_t,a_t,x_{t+1}))^2\right]
\right).
\]
Both the prior and posterior are parameterized by small multilayer perceptrons that output the mean and log-variance of the corresponding Gaussian distribution.

The recurrent state is updated by a GRU cell,
\[
h_{t+1}=f_\psi(h_t,a_t,z_t),
\]
where the GRU input is the concatenation of the action representation and latent transition state. During training on observed trajectories, the recurrent update uses a posterior latent sample $z_t^q \sim q_\phi(z_t\mid h_t,a_t,x_{t+1})$. During latent-space analysis, we use the prior mean.

The model predicts the next node and reward from the latent state and updated recurrent state. A categorical decoder outputs logits over all maze nodes,
\[
p_\theta(x_{t+1}\mid z_t,h_{t+1}),
\]
and a separate reward head outputs the probability of reward at the next node,
\[
p_\theta(r_{t+1}=1\mid z_t,h_{t+1}).
\]
The same decoder and reward head can be applied to the posterior path, for training against observed outcomes, or to the prior path, for prediction, action selection, value-map construction, and latent-space analyses.

\subsection{Training objective}

The model is trained online on trajectories produced by the current policy. For each generated bout, the model is updated using short temporal windows from that bout (window size = 20 steps). The total loss is
\[
\mathcal{L}
=
\mathcal{L}_{x}
+
\beta \mathcal{L}_{\mathrm{KL}}
+
\lambda_r \mathcal{L}_{r}
+
\lambda_s \mathcal{L}_{\mathrm{sparse}},
\]
where $\beta$ is warmed up during early training with its maximum value at $\beta_{max} = 10^{-3}$, along with $\lambda_r = 1$ and $\lambda_s = 0.1$.
The next-node loss is categorical cross-entropy,
\[
\mathcal{L}_{x}
=
-\log p_\theta(x_{t+1}\mid z_t^q,h_{t+1}^q),
\]
where
\[
z_t^q \sim q_\phi(z_t\mid h_t,a_t,x_{t+1}),
\qquad
h_{t+1}^q=f_\psi(h_t,a_t,z_t^q).
\]
The reward loss is binary cross-entropy,
\[
\mathcal{L}_{r}
=
-
\left[
r_{t+1}\log \hat r_{t+1}
+
(1-r_{t+1})\log(1-\hat r_{t+1})
\right],
\qquad
\hat r_{t+1}
=
p_\theta(r_{t+1}=1\mid z_t^q,h_{t+1}^q).
\]
Because reward events are sparse, the reward loss uses class balancing.

The KL term penalizes disagreement between the posterior inferred after observing the transition and the prior predicted before observing the transition:
\[
\mathcal{L}_{\mathrm{KL}}
=
D_{\mathrm{KL}}
\left[
q_\phi(z_t\mid h_t,a_t,x_{t+1})
\;\|\;
p_\theta(z_t\mid h_t,a_t,x_t)
\right].
\]
The sparsity term regularizes the posterior latent mean,
\[
\mathcal{L}_{\mathrm{sparse}}
=
\|\mu^q_\phi(h_t,a_t,x_{t+1})\|_1.
\]
In all main experiments, the same architecture, objective, optimizer settings, reward location, and online update procedure are used across agents; only the reward-driven switching probability $p$ is varied. All experiments were conducted on NVIDIA H100 HGX GPU nodes with 64-core CPUs. The reported experiments required approximately 8-10 hours.

\subsection{Expected information gain}

For each valid candidate action, the exploratory policy estimates expected information gain from the prior and posterior latent entropies. Given the current node $x_t$, recurrent state $h_t$, and candidate action $a$, the model first computes the prior
\[
p_\theta(z_t\mid h_t,a,x_t).
\]
Using the prior mean, the model predicts a distribution over possible next nodes. We then sample $K$ hypothetical next observations,
\[
\hat{x}_{t+1}^{(1)},\dots,\hat{x}_{t+1}^{(K)}
\sim
p_\theta(x_{t+1}\mid z_t,h_{t+1}),
\]
and for each sampled observation compute the corresponding posterior
\[
q_\phi(z_t\mid h_t,a,\hat{x}_{t+1}^{(k)}).
\]
The expected posterior entropy is approximated by Monte Carlo averaging:
\[
\frac{1}{K}
\sum_{k=1}^{K}
H\!\left[
q_\phi(z_t\mid h_t,a,\hat{x}_{t+1}^{(k)})
\right].
\]
The EIG score is
\[
\mathrm{EIG}(x_t,a)
=
H\!\left[
p_\theta(z_t\mid h_t,a,x_t)
\right]
-
\frac{1}{K}
\sum_{k=1}^{K}
H\!\left[
q_\phi(z_t\mid h_t,a,\hat{x}_{t+1}^{(k)})
\right].
\]
Actions in the exploratory regime are sampled with probability proportional to their nonnegative EIG scores across locally valid actions. If all valid scores are near zero, the policy falls back to a uniform distribution. The EIG score is used only for action selection and is not added to the training loss.

\subsection{Reward map and value propagation}

The reward-driven policy is built from the model's own prior reward predictions. Over a window of recent bouts ($n=50$), we average the predicted reward probability at each visited node to obtain a learned reward map,
\[
\hat{R}(x)
=
\frac{1}{C(x)}
\sum_{t:\,x_{t+1}=x}
p_\theta(r_{t+1}=1\mid z_t^p,h_{t+1}^p),
\]
where $C(x)$ is the number of recent visits to node $x$, and $z_t^p,h_{t+1}^p$ denote the prior latent state and corresponding predicted recurrent state.

We also construct an empirical transition graph from transitions observed in the same recent bouts. Value is propagated only over this observed graph using the soft backup
\[
V(x)
\leftarrow
\hat{R}(x)
+
\gamma T
\log
\sum_{x'\in\mathcal{N}_{\mathrm{obs}}(x)}
\exp\left(\frac{V(x')}{T}\right),
\]
where
\[
\mathcal{N}_{\mathrm{obs}}(x)
= \{x' : x\to x' \text{ was observed in recent bouts}\}.
\]
Here $\gamma = 0.8$ is the discount factor and $T=0.1$ is the softmax temperature. The value map is updated periodically (every $50$ bouts) from recent experience. In reward-driven mode, the agent selects the valid action leading to the neighboring node with highest learned value:
\[
a_t
=
\arg\max_{a\in\mathcal{A}(x_t)}
V(x_{t+1}^{a}),
\]
where $x_{t+1}^{a}$ denotes the deterministic next node reached by taking action $a$ from $x_t$.

\subsection{Mode switching}

The agent alternates between exploratory and reward-driven behavior. In exploratory mode, valid actions are sampled according to the EIG-based policy. At each exploratory step, the agent switches into reward-driven mode with probability $p$. Once reward-driven mode is active, the agent follows the value-map policy until it reaches the reward node. After reward delivery, reward-driven mode is disabled and the agent is forced to explore for $n=7$ steps.

The parameter $p$ is therefore a probability of entering reward-driven mode, not the fraction of reward-driven actions. Larger values of $p$ produce agents that are more likely to initiate reward-driven episodes, but the realized fraction of reward-driven actions also depends on how long the value-map policy takes to reach the reward.

\begin{figure}[t]
  \centering
  \includegraphics[width=1\linewidth]{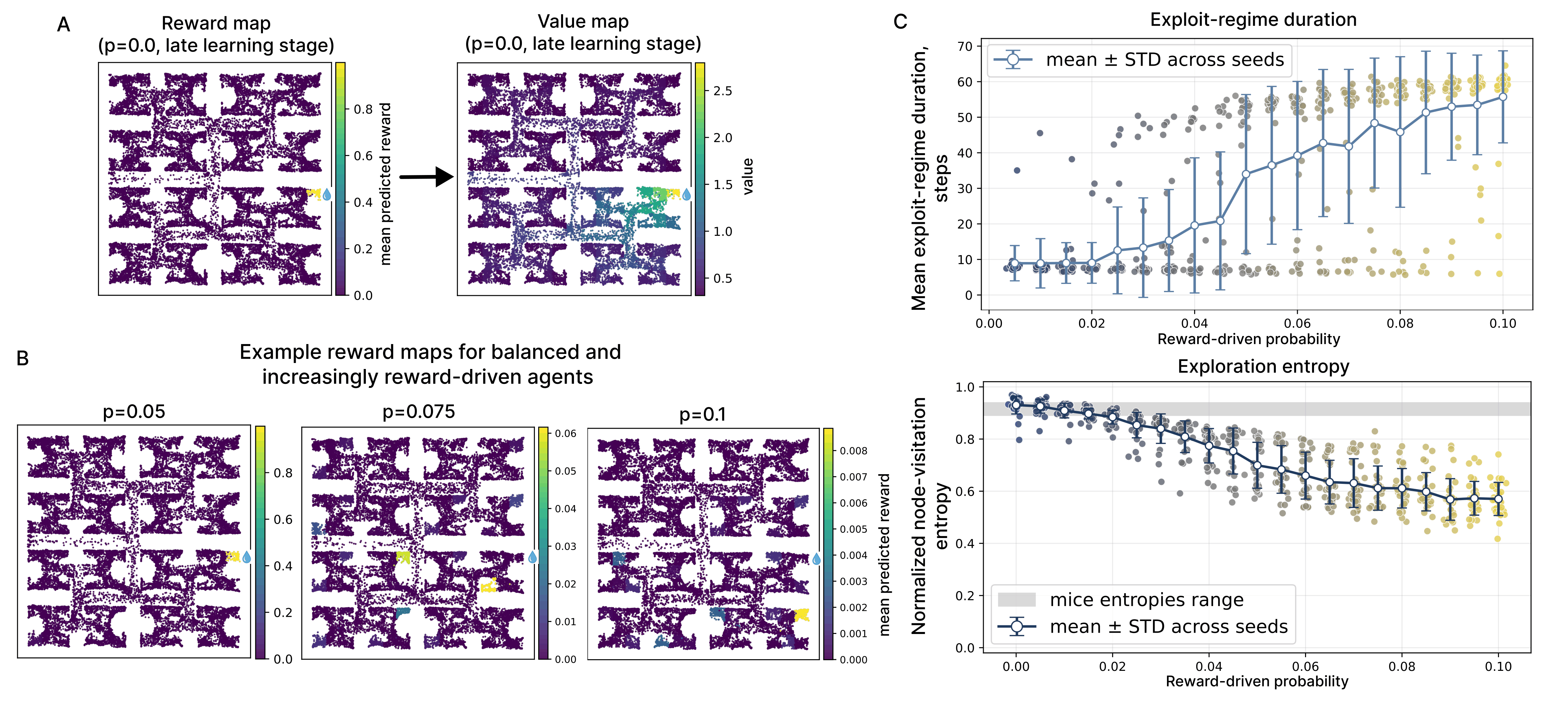}
  \caption{
\textbf{Additional analyses of reward-map learning and behavioral regimes.}
\textbf{A:} Transformation of the learned reward map into a value map.
\textbf{B:} Example learned reward maps from individual agents, shown without averaging across random seeds.
\textbf{C:} Relationship between reward-driven switching probability and behavioral regime. 
\textbf{Top:} Duration of reward-driven episodes as a function of the switching probability $p$. 
\textbf{Bottom:} Exploration entropy as a function of $p$, computed analogously to the mouse exploration measure in Fig.~\ref{fig:fig3}C. The gray band indicates the range of mouse visitation entropies, suggesting that mouse behavior approximately overlaps with agents using low reward-driven switching probabilities ($p=0$ to $p=0.025$). Individual points show single agents, with 30 agents per value of $p$; solid lines and error bars show the mean and standard deviation across agents.
}
  \label{fig:app2}
\end{figure}

\begin{figure}[t]
  \centering
  \includegraphics[width=1\linewidth]{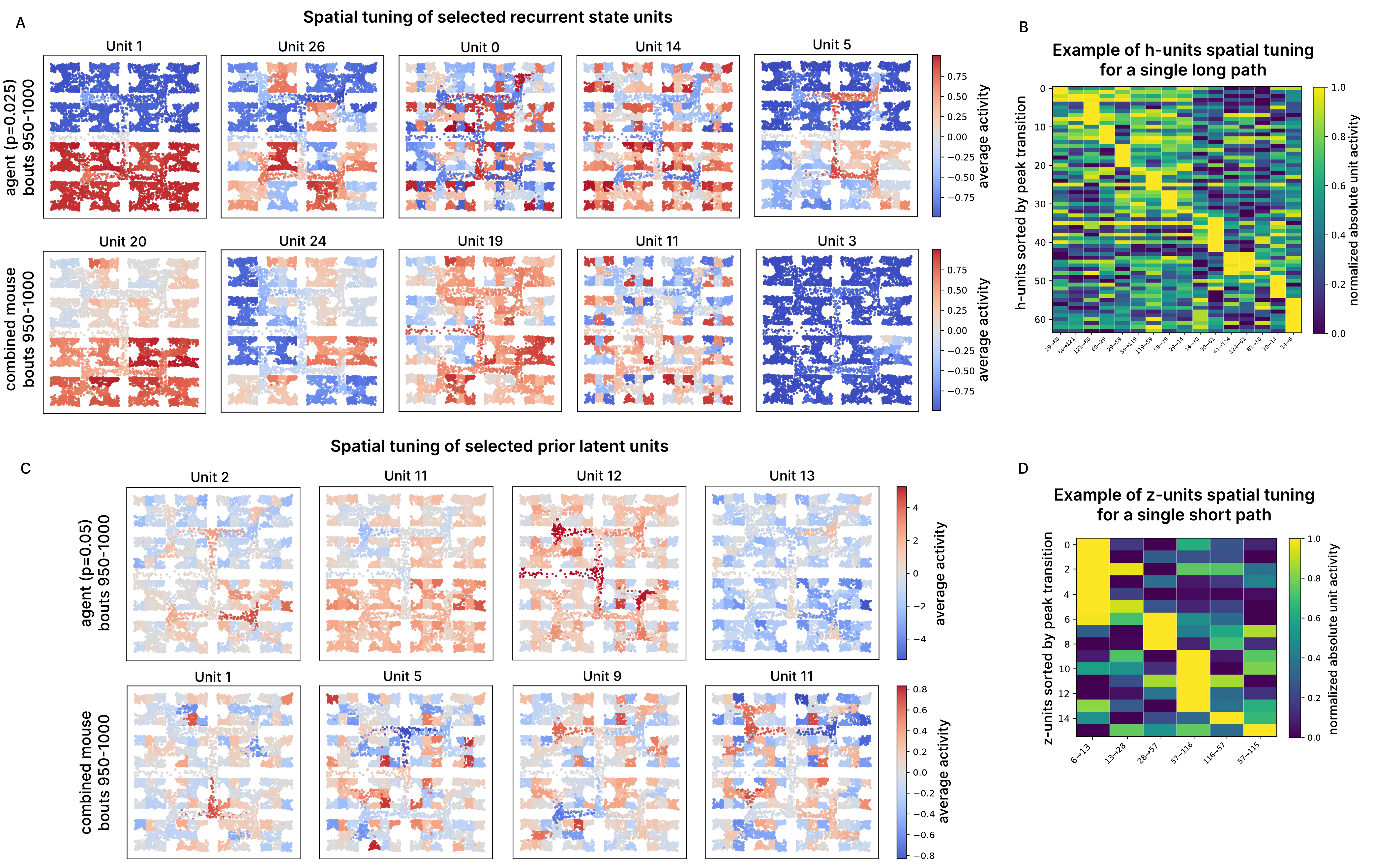}
  \caption{
\textbf{Single-unit spatial and path-transition tuning.}
\textbf{A:} Selected units from the recurrent state $h_t$ show spatial tuning across the maze. Examples include units tuned to broad maze regions, smaller subregions, terminal leaves, and specific paths. Similar tuning patterns are observed in a long-trained agent and in a combined-mouse model.
\textbf{B:} Path-transition tuning of $h_t$ units along an example trajectory. Columns correspond to successive transitions along the selected path, and rows correspond to units sorted by the transition at which their activity peaks. Activity is averaged over transition occurrences in late bouts and normalized within each unit.
\textbf{C:} Same as panel A, but for units of the predictive latent state $z_t$.
\textbf{D:} Path-transition tuning of $z_t$ units along an example water-port path. Rows and columns are organized as in panel B, showing that subsets of latent units peak at different transitions along the path.
}
  \label{fig:app3}
\end{figure}

\section{Spatial tuning of model units}
\label{app:unit_tuning}

The main analyses focus on population-level geometry of the prior latent space. We additionally examined whether individual units in the recurrent state $h_t$ and predictive latent state $z_t$ develop spatially interpretable tuning. These analyses are not used to define the main geometry metrics, but provide a complementary unit-level view of the learned representation.

We analyzed units from two settings: a long-trained agent after late training bouts, and a combined-mouse model trained on mouse trajectories. For the combined-mouse model, bouts from individual mice were interleaved while preserving relative learning order: first bouts from all mice were processed before second bouts, followed by later bouts in the same way. This produces a single model trained on the aggregate distribution of mouse trajectories while approximately preserving the temporal progression of experience across animals.

We first visualized spatial tuning across the full maze (Fig.~\ref{fig:app3}A,C). For each unit, we averaged its activity over visits to each node and projected the resulting activity values back onto the binary-tree layout. Units in both $h_t$ and $z_t$ showed structured tuning at multiple spatial scales, including broad subtrees, smaller local regions, terminal leaves, and specific paths. The recurrent state $h_t$ summarizes the sequence of transitions leading up to the current step, while the predictive latent state $z_t$ represents the current action-conditioned transition. Thus, tuning in $h_t$ can reflect accumulated trajectory context, whereas tuning in $z_t$ can reflect the transition currently being predicted.

We next examined tuning along selected paths through the maze (Fig.~\ref{fig:app3}B,D). For a chosen sequence of nodes, we treated each consecutive pair as a path transition. For each transition, we averaged the absolute activity of each unit across all occurrences of that transition in late bouts. We then normalized each unit's activity across transitions and sorted units by the transition at which they reached peak activity. The resulting heatmaps show whether different units are preferentially active at different positions along a path. These results suggest that the population-level geometry described in the main text is accompanied by interpretable unit-level tuning, including tuning to broad maze regions, local branches, leaves, and specific transition sequences. We do not interpret these units as direct neural analogues, but as model-based examples of the spatial and transition-dependent tuning that can emerge from predictive learning on navigation trajectories.


\end{document}